\documentclass[twocolumn]{aastex631}
\usepackage{amsmath}

\newcommand{\lya}{Ly$\alpha$}

\newcommand{\g}{{\rm\thinspace g}}

\newcommand{\s}{{\rm\thinspace s}}

\newcommand{\fesc}{$f_{\rm esc}$}

\newcommand{\G}{{\rm G}}
\newcommand{\K}{{\rm K}}

\begin{document}

\title{Systematic Bias in Ionizing Radiation Escape Fraction Measurements from Foreground Large-Scale Structures}

\author[0000-0002-9136-8876]{Claudia M. Scarlata}
\affiliation{Minnesota Institute for Astrophysics, School of Physics and Astronomy, University of Minnesota, 316 Church Street SE, Minneapolis, MN 55455, USA}
\author{W.~Hu}
\affiliation{Minnesota Institute for Astrophysics, School of Physics and Astronomy, University of Minnesota, 316 Church Street SE, Minneapolis, MN 55455, USA}
\author[0000-0001-8587-218X]{Matthew J. Hayes}
\affiliation{Stockholm University, Department of Astronomy and Oskar Klein Centre for Cosmoparticle Physics, AlbaNova University Centre, SE-10691, Stockholm, Sweden}
\author{S.~Taamoli}
\affiliation{Physics and Astronomy Department, University of California Riverside}
\author[0000-0002-0101-336X]{Ali A. Khostovan}
\affiliation{Department of Physics and Astronomy, University of Kentucky, 505 Rose Street, Lexington, KY 40506, USA}
\author{C. M.~Casey}
\affiliation{Department of Physics, University of California, Santa Barbara, CA 93106}
\affiliation{Department of Astronomy, the University of Texas at Austin, 2515 Speedway Blvd Stop C1400, Austin, TX 78712}
\affiliation{Cosmic DAWN Center}
\author[0000-0002-9382-9832]{Andreas L. Faisst}
\affiliation{Caltech/IPAC, 1200 E. California Blvd. Pasadena, CA 91125, USA}
\author[0000-0001-9187-3605]{Jeyhan S. Kartaltepe}
\affiliation{Laboratory for Multiwavelength Astrophysics, School of Physics and Astronomy, Rochester Institute of Technology, 84 Lomb Memorial Drive, Rochester, NY 14623, USA}
\author[0000-0001-8792-3091]{Yu-Heng Lin}
\affiliation{Caltech/IPAC, 1200 E. California Blvd. Pasadena, CA 91125, USA}
\author[0000-0001-7116-9303]{M. Salvato}
\affiliation{Max-Planck Institute for Extraterrestrial Physics, Giessenbachstrasse1, 85748, Garching Germany}
\author[0000-0002-9946-4731]{Marc Rafelski}
\affiliation{Space Telescope Science Institute, 
3700 San Martin Drive, 
Baltimore, MD, 21218 USA}
\affiliation{Department of Physics and Astronomy, Johns Hopkins University, Baltimore, MD 21218,USA}



\begin{abstract}
We investigate the relationship between the \lya\ forest transmission in the intergalactic medium (IGM) and the environmental density of galaxies, focusing on its implications for the measurement of ionizing radiation escape fractions. 
Using a sample of 268 spectroscopically confirmed background galaxies at $2.7<z<3.0$ and a galaxy density map at $z\approx 2.5$ within the COSMOS field, we measure the \lya\ transmission photometrically, leveraging the multiwavelength data available from the COSMOS2020 catalog. Our results reveal a weak but statistically significant positive correlation between \lya\ optical depth and galaxy density contrast, suggesting that overdense regions are enriched in neutral gas, which could bias escape fraction measurements. This emphasizes the need to account for the large-scale structure of the IGM in analyses of ionizing radiation escape fractions, and highlights the advantages of a photometric approach for increasing the number of sampled lines of sight across large fields. The photometric redshifts provided by upcoming all-sky surveys, such as Euclid, will make it possible to account for this bias, which can also be minimized by using  fields separated in the sky by many degrees.
\end{abstract}

\section{Introduction} \label{sec:intro}
New observations from the James Webb Space Telescope are revolutionizing our understanding of the galaxies that populate the Universe during the epoch of reionization \citep[i.e., at $z\gtrsim 6$][]{Williams2023,lin_empirical_2024, Finkelstein2024}. Not only  are we able to routinely identify and spectroscopically confirm objects at these epochs, but we are now able to study their physical properties in detail \citep[see, e.g.,][for a recent review]{Adamo2024}. These studies are  showing that galaxies at $z\gtrsim 6$ share many of the properties of local galaxies with high escape fraction of ionizing radiation \citep{lin_empirical_2024,Harikane2024,Hayes2024}, possibly confirming the role of star-formation in the reionization of the Universe, although the role of active galactic nuclei is still debated \citep[e.g.,][]{Madau2024}. This conclusion, however, is weakened by our ignorance of how much ionizing radiation from $z>6$ galaxies escapes into the inter galactic medium (IGM). 

There are no observations that can measure the escape of ionizing radiation at $z\gtrsim 4.5$, as the diffuse neutral hydrogen in the IGM prevents leaked ionizing photons from reaching the observer. Accordingly, studies of escape fraction have  focused on $z\lesssim 4$, with mixed success \citep[e.g.,][]{Citro2024,Vanzella2018}. 

One of the primary challenges faced by escape fraction studies at $z \gtrsim 1.6$ is the uncertainty in correcting for the IGM absorption  specific to the line of sight to each galaxy \citep{steidel_keck_2018}. To overcome this, many high-redshift studies report average escape fractions, operating under the assumption that averaging across multiple sight lines allows for the use of the average attenuation factor measured from large numbers of QSO independent sight lines  \citep{fan_constraining_2006, becker_evidence_2015, becker2018}.

\begin{figure*}[ht!]
\begin{center}
\includegraphics[width=19cm]{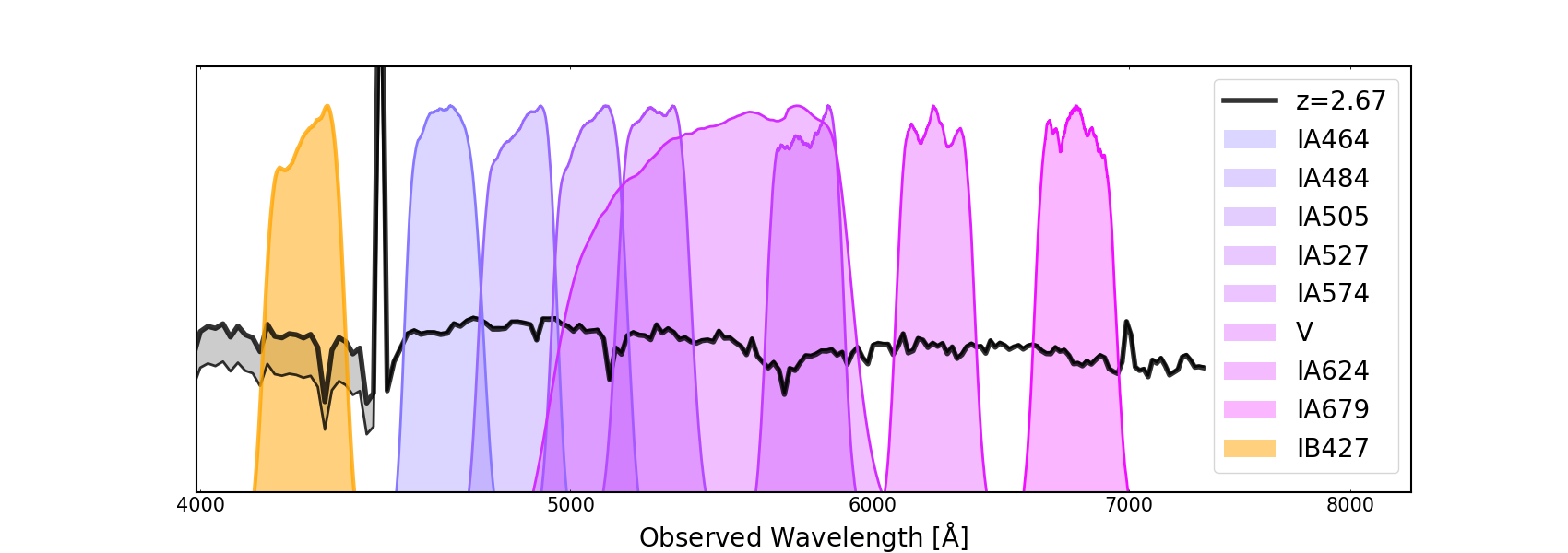}
\caption{COSMOS filter transmission curves used in this work, overlaid to the spectral template of a star-forming galaxy at $z=2.67$ see text for details. The thick and thin black lines show the spectrum before and after absorption by the \lya\ forest. The IB427 filter samples the portion of the absorption due to material in the   $ 2.4 \lesssim z \lesssim 2.6$ redshift range.   }\label{fig:method}
\end{center}
\end{figure*}

\citet{Steidel2018} demonstrated that the uncertainty related to the unknown effects of the intergalactic medium (IGM) can be reduced to less than 10\% at $z\approx 3$ by averaging over a large number ($\gtrsim 60$) of lines of sight. However, these lines of sight must be truly independent for the statistical argument to hold. In cases where galaxies are observed over small patches of sky, this assumption may not be valid due to the clustering of the underlying matter density. For galaxies at 
$z\approx 3$, LyC photons are absorbed both `locally' by optically thick Lyman limit systems (LLSs) and by the \lya\ forest at lower redshifts. Given that the mean free path of a LyC photon emitted at $z=3$ is relatively large \citep[on the order of 100 pMpc,][]{Becker2021}, a significant fraction (approximately 40\%) of the absorption results from the cumulative effect of the \lya\  forest at a redshift of around 2 for the galaxy in question. This latter component of the IGM attenuation is what can potentially introduce a correlation between the IGM correction and the large scale structure in the foreground. 

Various studies suggest that the \lya\ forest absorption correlates on scales up to tens of comoving Mpc \citep[e.g.,][]{Liske2000, Slosar2011,Mawatari2023}. At $z \approx 3$, angular scales of 10 arcminutes  correspond to approximately 20~cMpc implying that nearby sightlines  are not completely independent.  Existing measurements of LyC escape fraction at $z\sim 3$ are confined to a handful of  well--studied and relatively small fields \citep{Naidu2017,fletcher_lyman_2019,Saxena2022}, and are  therefore affected by this limitation. Even the `large' CANDELS fields only span linear scales of at most 18\,cMpc, at $z=3$. 
Applying an  ``all-sky-averaged" IGM transmission to small fields may not  appropriate.

A few studies in the literature have observationally explored this effect. \citet{mukae_cosmic_2017} investigated \lya\ forest absorption at $z\approx 2.5$ along the lines of sight to nine quasars in the COSMOS field. Using galaxy densities estimated within $5\times 25$ pMpc cylinders, they detected a weak correlation between the \lya\ forest attenuation and galaxy density. Similarly, \citet{Liang2021} found comparable results by cross-correlating galaxy density fields, traced by \lya-emitters, with \lya\ forest attenuation measured against background quasars from the SDSS Baryon Oscillation Spectroscopic Survey (BOSS) database. A common limitation of these studies is their reliance on \lya\ forest attenuation measurements from the spectra of bright quasars, which restricts the analysis to a limited number of lines of sight. In this work, we adopt a different approach by using photometric data from galaxies rather than spectroscopic data from quasars. This method allows for a significantly larger number of lines of sight within a field, enhancing the sampling of the underlying large-scale structure, and emphasizing smaller transverse spatial scales.

In this paper we apply this approach to the COSMOS field \citep{Scoville2007}, where the wealth of ancillary data (Section~\ref{sec:data}) allows an accurate measurement of both the \lya\ forest attenuation (Section~\ref{sec:opticaldepth}) and the galaxy density field (Section~\ref{sec:density}). In Section~\ref{subsec:results} we present  our results and discuss how the foreground LSS affects the measurement of the LyC escape fraction.  In what follows, we assume a cosmology of $\{H_0,\Omega_\mathrm{M},\Omega_\Lambda\} = \{70~\mathrm{km~s}^{-1}~\mathrm{Mpc}^{-1},0.3, 0.7\}$, and the AB magnitude system \citep{Oke1990}. 

\section{The strategy} \label{sec:data}
\subsection{Theory and available filters} 
As a galaxy's radiation traverses the IGM toward the observer, a series of absorption lines (typically referred to as the \lya\ forest, even though they are due to both hydrogen and metals) populate the spectrum blueward of the \lya\ emission line observed at ((1+$z_G$)$\times 1216$\AA, where $z_G$ is the redshift of the galaxy).
For a source at redshift $z_G$, the LyC radiation that is not absorbed by a Lyman limit system in the galaxy's local environment can travel almost freely until it resonates with the \lya\ line of neutral hydrogen at redshift $z_{\rm forest} = \frac{\lambda_{<912}}{1216}(1+z_G) - 1$. Resonant absorption in the higher order Lyman series is also possible, but the \lya\ cross section is 5.3 times higher than Ly$\beta$ and thereby provides the large majority of the opacity. 
We use photometric and spectroscopic data from the COSMOS survey to quantify the \lya\ forest absorption at $z_{forest}\approx 2.5$ using a sample of background galaxies with accurate spectroscopic redshifts. 

In Figure~\ref{fig:method}, we illustrate the concept behind the measurement. The thick black line represents a spectral template of a  10~Myr old star-forming galaxy with an exponentially declining star-formation rate, and timescale of 10Myrs at $z_G=2.67$ before its light is attenuated by the intervening IGM (computed using the average value for $z=2.67$). After passing through the IGM, the galaxy's light is diminished by the \lya\ forest, as shown by the spectrum given by the thin line. The gray shaded  area emphasizes the difference between the attenuated and unattenuated spectrum. 
The absorption due to the \lya\ forest  can be measured using the combination of photometric bands available in COSMOS  and shown in the Figure. Specifically, the COSMOS IB427 medium band filter (covering the wavelength range between 4170--4370\AA) isolates \lya\ forest absorption from gas at $\langle z \rangle = 2.5$, between $z_{min}=2.41$ and $z_{max}=2.59$ (corresponding to a scale of $\approx 40$~pMpcs). Accordingly, for background galaxies at $z_G>2.65$, the comparison between the intrinsic and observed flux in the COSMOS IB427 filter provides an estimate of the IGM attenuation due to gas at $\langle z \rangle = 2.5$.

In the following sections, we explain the selection of the background sources,  the calculation of the \lya\ forest attenuation using the available photometric bands, and the calculation of the density contrast.

\begin{figure}[ht!]
\includegraphics[width=8.5cm]{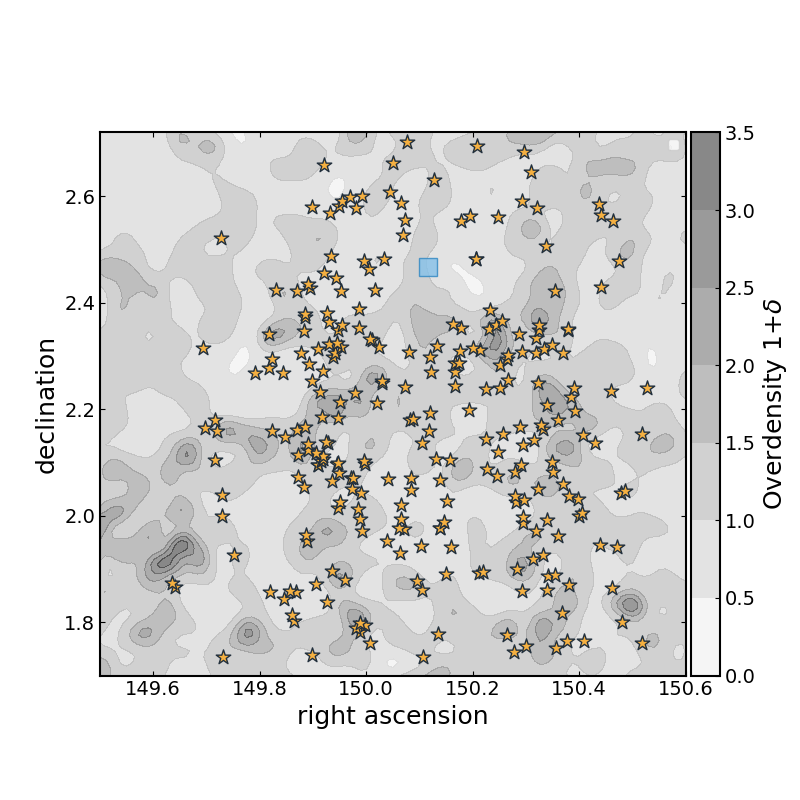}
\caption{Map of the overdensity parameter (1$+\delta$) at $\langle z\rangle =2.5$,  with the locations of the selected galaxies represented by yellow stars. The turquoise square shows the size of one Wide Field Camera~3 field of view, as reference. 
\label{fig:map}
}
\end{figure}

\subsection{Background galaxy sample}\label{subsec:bkg}
To isolate the \lya\ forest transmission, and minimize the modeling uncertainty that could be introduced by contamination from \lya\ emission from the background galaxies, we limit the analysis to background objects in the $2.65<z_{bkg}<3$. The low redshift limit ensures that the IB427 photometry is not contaminated by possible \lya\ emission of the background galaxy, while the high redshift cutoff avoids the Ly$\beta$ absorption.
We use the latest compilation of spectroscopic redshifts in COSMOS, compiled by \citet{cosmos_z} as the parent catalog. This  catalog includes $\approx 10^5$ unique sources with spectroscopic redshifts either published in the astronomical literature or private to the COSMOS collaboration. From the compilation, we used the astrometry-corrected coordinates and the homogenized quality flags.  We only include spectroscopic redshifts  with quality assessment flag  equal to either 3 or 4.  This choice limits the selection to only reliable sources with a redshift confidence level $> 95$\%.  The reader is referred to Section~3.1 of \citet[][]{cosmos_z} for the details of how the quality flags are defined and homogenized in the compilation. When sources are observed in multiple surveys, we keep the preferred redshift characterized by the {\tt\string Priority=1} assessment flag. To ensure consistent photometric coverage in all bands, we  limit to the central COSMOS area, with  $149.48< $R.A.$< 150.55$ and  $1.7<$ Dec $< 2.72$. 
Finally, we excluded sources with photometric and spectroscopic redshifts that differed by more than 0.2, as well as objects that were identified as AGN in the literature \citep[e.g.,][]{Laloux2023}.  We note that the spectra cannot be used for the calculation of the \lya\ optical depth as they are not of sufficient quality and/or do not cover the full wavelength range required for the analysis. 

For each galaxy in the spectroscopic sample we extracted the photometric measurements tabulated in the Farmer$+$Lephare COSMOS2020 catalog \citep{weaver_cosmos2020_2022}, and we retain in the sample only galaxies with a signal-to-noise ratio greater than three in the IB427 filter, reducing the sample to 268 galaxies. 

\subsection{Galaxy environmental density maps}\label{sec:density}
We employ the galaxy density maps (specifically, the map of 1$+\delta$, where $\delta = \frac{\rho-\rho_m}{\rho_m}$,  and  $\rho_m$ is the mean density of galaxies in the redshift slice) from \citet{taamoli_large-scale_2024}, who reconstructed the underlying galaxy  density field using the full photometric redshift probability distribution functions for galaxies in the COSMOS2020 catalog. The maps were computed in narrow redshift ranges using a weighted kernel density estimation approach, and correcting for edge effects and masked regions. In this work, we consider the stacked density map obtained  within redshift range covered by the IB427 filter in \lya, i.e., $2.4<z<2.6$. The density map is shown in Figure~\ref{fig:map}, where we also show the position of the background galaxies used for the measurement of the \lya\ forest transmission. 

For each background galaxy, we extract the corresponding value of 1$+\delta$ along its line of sight, by computing a simple linear interpolation of the map at the specific position. We estimate the error associated with this overdensity by taking the standard deviation of the map values in a circle of 2.5 arcminute radius (corresponding to $\approx 4$cMpc at $z=2.5$).

\subsection{Optical Depth Calculation}\label{sec:opticaldepth}
We compute the average \lya\ transmission over the $2.4-2.6$ redshift range along the line of sight to each galaxy from the ratio between the observed  ($f_{observed}$)  and intrinsic   galaxy flux in the IB427 filter ($f_{model}$), i.e., $ \langle T_{Ly\alpha} \rangle _{z=2.5} = f_{observed}/f_{model} $. The average \lya\  transmission is then  used to compute the effective \lya\ optical depth defined as:

\begin{equation}
    \tau_{Ly\alpha} = -\ln{\langle T_{Ly\alpha} \rangle _{z=2.5}}
\end{equation}

\begin{figure}[ht!]

\includegraphics[width=8.5cm]{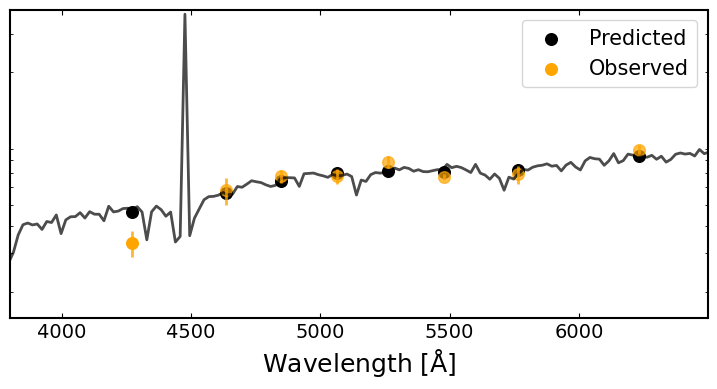}
\caption{Example SED modeling on a $z=2.78$  galaxy. The orange dots represent the measured fluxes in the COSMOS2020 catalog. The black dots are computed using the best-fit spectrum (shown in black). 
\label{fig:example}}
\end{figure}

To estimate $f_{model}$, we fit stellar population synthesis models to the background galaxies' photometric measurements, limiting the fit to only bands  sampling the spectral energy distribution between 1216\AA\ and 2000\AA\ in the rest frame of the galaxy, see Figure~\ref{fig:example}. We used the Bayesian Analysis of Galaxies for Physical Inference and Parameter Estimation code \citep[BAGPIPES][]{Carnall2018} to model the galaxies' spectral energy distribution. Because of the uncertain impact of \lya\ radiative transport in the ISM of galaxies, we exclude photometric bands that can potentially be contaminated by \lya\ emission from the SED fit. Accordingly, depending on the specific redshift, we use either 8 or 7 bands for each galaxy. Before fitting the SED, we corrected the photometric measurements for Milky Way extinction, using the \citep{Gordon2023} extinction law with a total-to-selective extinction $R_V=3.1$. Additionally, the fluxes in the individual photometric bands were corrected using the magnitude offsets computed by the EAZY code \citep[see details in ][]{weaver_cosmos2020_2022}.

For the BAGPIPES fit, we use \citet{bruzual_stellar_2003} stellar population synthesis models with an exponential star formation history and a Kroupa IMF \citet{kroupa_variation_2001}. We vary the stellar ages, metallicities, and we include both nebular emission and dust attenuation (using the \citet{calzetti_dust_2000} extinction law).  We checked that considering a different extinction law, \citep{reddy_connection_2016}, did not change the paper conclusions which are based on a relative comparison of observed transmissions among different lines of sight. During the fit, the redshift was kept fixed at the galaxy spectroscopic redshift.  

Figure~\ref{fig:example} shows an example of the modeling for an observed galaxy at $z=2.78$. The black line shows the intrinsic best-fit model, i.e., the model that does not include the \lya\ forest attenuation. We used this spectrum to compute $f_{model}$ within the IB427 passband and compared this value to $f_{observed}$ to compute $\tau_{Ly\alpha}$, according to Equation~1.  
We do not apply a correction for metal absorption lines. This correction, however, is negligible, being of the order of $\Delta \tau \approx$0.0245 \citep{kirkman_h_2005}.

\section{Results and Discussion}\label{subsec:results}
Before we investigate the correlation between the \lya\ forest optical depth ($\tau_{Ly\alpha}$) and the galaxy overdensity we compute the average value $\tau_{Ly\alpha}$ including all lines of sight. We find that $\tau_{Ly\alpha} = 0.33^{+0.18}_{-0.21}$, a value somewhat higher than the average optical depth at similar redshift estimated from QSO lines of sight \citep[e.g., $\tau_{Ly\alpha}^{\rm QSO} \approx 0.2$,][]{ kirkman_h_2005, monzon_effective_2020}. One possibility is that the specific lines of sight we have identified (i.e., those corresponding to $z\approx 3$ Lyman Break Galaxies) are more attenuated than the average line of sight as galaxies selected to be spectroscopically followed up are typically identified on the basis of a strong Lyman break feature. 
\begin{figure*}[ht!]
\includegraphics[width=8.cm]{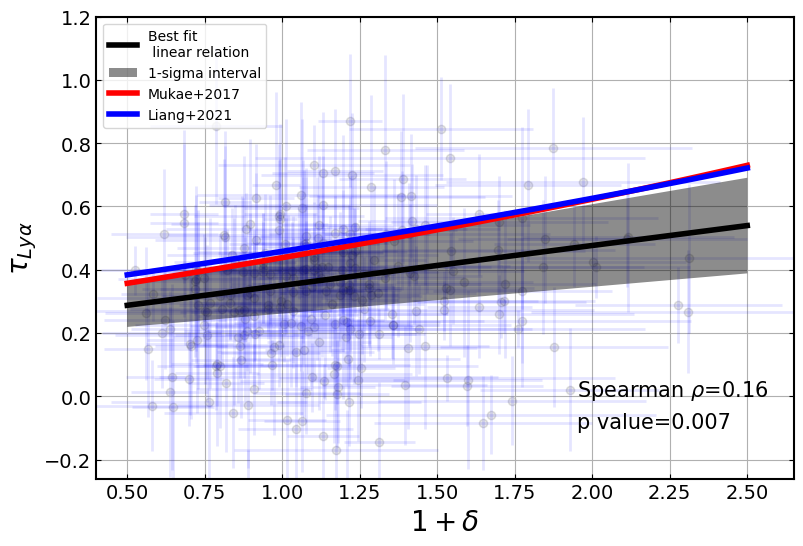}
\includegraphics[width=9cm]{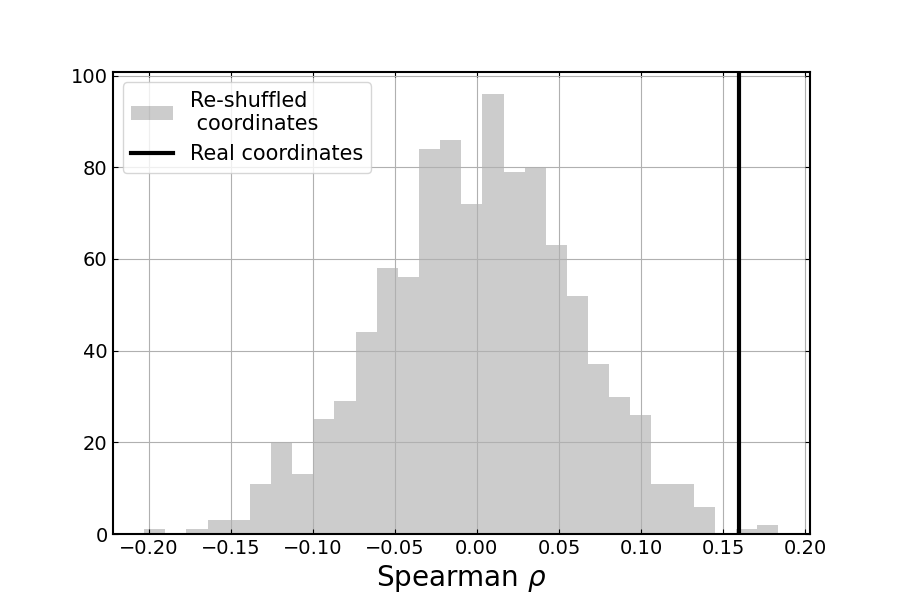}
\caption{(Left) \lya\ optical depth as a function of galaxy overdensity for the spectroscopically confirmed galaxies in COSMOS (gray points) and best linear fit with corresponding uncertainty (black line and gray areas). The correlation is weak but significant and agrees with  previous studies from the literature at similar redshifts (red and blue curves for \citet{mukae_cosmic_2017} and \citet{Liang2021}, respectively). (Right) Histogram of the Spearman $\rho$ coefficient computed for 1,000 random realizations of  the background galaxies' coordinates. The vertical  line shows the Spearman coefficient computed on the galaxies' observed position.
\label{fig:correlation}}
\end{figure*}

\subsection{The \lya\ optical depth and foreground galaxy density}\label{sect:res:tau_vs_dens}
The main result of our analysis is presented in Figure~\ref{fig:correlation} where we show the $\langle z \rangle=2.5$ effective \lya\ forest optical depth as a function of the density enhancement. The data show that there is a weak correlation between the \lya\ optical depth and the density field, with lines of sight passing through overdense regions being on average less transparent to \lya\ photons than lines of sight passing through underdense regions. 

We quantify the strength of the correlation using the Spearman's ($\rho_S$) correlation coefficient, based on the ranks of the data. We find that $\rho_S=0.16$ (with a p-value of 0.007), indicating that the correlation is weak, but significant. 
We also tested the reality of the correlation by shuffling the coordinates of the background galaxies, and recomputing the correlation coefficient for each realization. 
Because each background galaxy provides the attenuation along a specific sightline in the large scale structure at $z\approx2.5$, shuffling the coordinates should erase any correlation. The results are shown in Figure~\ref{fig:correlation}, where we plot the distribution of the Spearman's $\rho_S$ correlation coefficient between the \lya\ forest attenuation and the galaxy density contrast for the 1,000 realizations of the re-shuffled galaxy coordinates. The vertical line shows the value measured for the real galaxy positions. This figure clearly demonstrates that the observed correlation has a very low probability of rising by chance. The best-fit linear relation to the data, shown in black with the associated scatter, is:

\begin{equation}
\tau_{Ly\alpha} = (0.12 \pm 0.04)\times (1+\delta) + 0.22 \pm 0.05.
\end{equation}

The observed trend suggests that at $z\sim 2.5$, overdense environments are more enriched in neutral gas than less dense regions. This finding aligns with the observation that, at these redshifts, the average star formation rate (SFR) of galaxies is higher in denser environments-- a trend that is reversed in the local Universe \citep[e.g.,][]{lemaux_vimos_2022,taamoli_large-scale_2024}, and supports the idea that the increased SFR may be  due to higher level of gas accretion in overdense environments.

A few studies have approached this problem in the literature measuring the \lya\ forest opacity using the spectra of bright background QSOs. \citet{mukae_cosmic_2017} applied this technique to 16 QSOs in the COSMOS field, where they used galaxies' photometric redshifts to estimate the environmental densities within 25~pMpc-high\footnote{In the redshift direction.} cylinders, corresponding to  $\Delta(z)=0.0875$, approximately half of the length of the redshift slice covered by the IB427 filter. \citet{Liang2021} used 64 QSO lines and constructed the galaxy density field using \lya\ emitters (LAEs) photometrically identified at $z=2.18$ over a narrow redshift range\footnote{LAEs may or may not be a good tracer of the underlying density field \citep[e.g.,][]{Shi2019,Toshikawa2016} as the escape of \lya\ radiation from galaxies is a complex function of galaxies properties. }. For the measurement of $\tau_{Ly\alpha}$ they averaged the IGM transmission over $\approx 6.7$~pMpc. Similar to our results, both studies find a weak but significant correlation between the \lya\ forest transmission and galaxy overdensity. The \citet{mukae_cosmic_2017} and \citet{Liang2021} relations are shown in Figure~\ref{fig:correlation}. In both  works, they presented the results in terms of the \lya\ forest fluctuation, i.e., the fluctuation of the \lya\ forest transmission with respect to the cosmic \lya\ forest mean transmission, $\delta_T$.  For the mean cosmic transmission, \citet{mukae_cosmic_2017} and \citet{Liang2021} adopt the value estimated by \citet{faucher-giguere_direct_2008}, $T_{cosmic} =0.78$, corresponding to a $\tau_{cosmic}\approx 0.25$. Converting from $\delta_{T}$ to $\tau$ can be done with simple algebra, i.e., $\tau = -\ln(\delta_T+1)+0.25$. 

As can be seen from the comparison in Figure~\ref{fig:correlation} our result agrees with these previous studies in terms of the observed slope of the correlations, but is slightly shifted toward lower values of optical depths,  although within observational errors. If real, this shift is hard to understand. From QSO absorption line studies, the cosmic average \lya\ forest transmission of the IGM at $z=2.5$ is $\tau_{Ly\alpha}^{QSO} \approx 0.2$, varying relatively little between redshift 2.2 and 2.6. We already noted that the average value we find in our analysis is larger than this cosmic average, and we suggested the possibility that this difference is due  to the choice of background sources (galaxies instead of QSOs). The two studies mentioned above, however, target bright QSOs not galaxies. It is also unlikely that this difference is due to the specific lines of sight being considered, as the \citet{mukae_cosmic_2017} measurement is performed in the COSMOS field over a very similar redshift range as used here, making it even more difficult to reconcile the different values. 

Finally, Figure~\ref{fig:correlation} shows that although the slope derived in our work is somewhat shallower than the slopes of the relations reported in  \citet{mukae_cosmic_2017} and \citet{Liang2021} the curves are within the one $\sigma$ error.  A possible shallower trend could be due to variations in the volumes used to compute galactic environmental densities and optical depths, with our analysis employing the longest cylinders. Averaging over larger volumes likely weakens the correlation, which however is still found to be significant over the 40~cMpc scale  ($\approx 200$\AA) used here. 

\subsection{Impact on \fesc\ measurements}
We now focus on how the dependency of $\tau_{Ly\alpha}$ on the galaxies'  environmental density affects  the measurement of the escape fraction ($f_{esc}$) of ionizing radiation. 

 The escape fraction of ionizing radiation is computed from a galaxy's flux below 912\AA. For galaxies below redshift $\approx 1.5$, ionizing photons that escape from the galaxy's interstellar and circumgalactic medium can travel freely through the intergalactic medium and reach the observer. At higher redshifts, however, measuring  \fesc\ requires  a correction for IGM absorption \citep[$e^{\tau_{IGM}}$, see, e.g.,][]{Siana2015}, where $\tau_IGM$ is known only statistically from  absorption line studies of QSOs distributed over the full sky \cite[][]{Inoue2014,Prochaska2019}. 
$\tau_{IGM}$ can be due to two main mechanisms \citep{Madau1995}: As photons travel, they can photoionize neutral hydrogen in the vicinity of the galaxy (most notably in  Lyman Limit Systems, LLSs) or they can be scattered (in the \lya\ transition) by neutral gas much further away from the galaxy, i.e., the \lya\  component studied here. 

The number of LLSs seen by the traveling photons introduces an uncertainty to the absorption that was studied by \citet{Steidel2018}. This uncertainty is random and can be mitigated by averaging measurements across a large number of galaxies. In contrast, the attenuation due to \lya\ scattering can add a systematic uncertainty to the IGM correction to \fesc, as we found that $\tau_{Lya}$ correlates with the intervening large scale structure. As shown in Figure~\ref{fig:map}, overdense and underdense regions can extend across scales of several arcminutes (with the turquoise square representing the size of a Wide Field Camera 3 field of view). Galaxies located behind these regions will systematically experience greater (for overdense) or lesser (for underdense) IGM attenuation, requiring a correspondingly larger or smaller correction when measuring their escape fraction of ionizing radiation.

\begin{figure}[h]
\includegraphics[width=8cm]{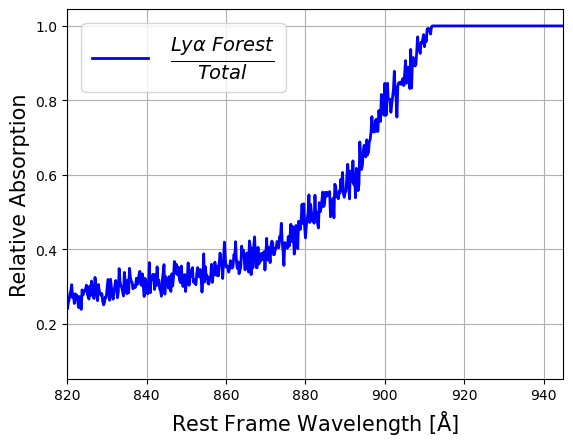}
\caption{Simulation of the IGM absorption as a function of wavelength due to a spatially unclustered universe. The blue curve shows the relative absorption of the \lya\ forest compared to the total IGM absorption (photoionization $+$ forest). At $z=3.6$, considered in this simulation, the low redshift intervening forest contributes between 90 and 30\% of the total absorption of the ionizing radiation at wavelengths of 910 and 840\AA, respectively.   
\label{fig:relative}}
\end{figure}

To assess the relative contributions of photoionization and \lya\  forest absorption to IGM attenuation, we simulated the absorption of ionizing radiation following the method outlined in \citet{inoue_updated_2014}. This simulation incorporates their parameterization of redshift- and column-density distributions, which were derived from widely separated QSO sight lines and thus assume a spatially uncorrelated universe. These distributions do not include an excess of absorbers associated with the circumgalactic medium \citep[e.g., ][]{rudie_column_2013}. Note, however, that \citet{hayes_spectral_2021} find no need for this additional absorption around galaxies based on the evolution of the average \lya\ profile (an effect that can only occur proximately) with redshift.  We generated 500 sight lines to 
$z=3.6$, computing both the total IGM absorption (including photoionization and \lya\ forest absorption) and the absorption due to the \lya\ forest alone. In Figure~\ref{fig:relative}, we show the relative contribution of the \lya\ forest absorption to the total absorption by the IGM. The \lya\ forest alone accounts for a substantial fraction ranging from 30\% to 90\% of the total attenuation at wavelengths between 910\AA\ and 840\AA. This result highlights the significant role that large-scale structure plays in modulating the observed ionizing radiation, reinforcing the need to account for these environmental effects when interpreting escape fraction measurements at redshifts above $\approx$ 1.5.

In Section~\ref{sect:res:tau_vs_dens} we considered the possibility that the slope of the $\tau_{Lya}-\delta$ relation depends on the wavelength interval ($\Delta \lambda$) over which it is computed.  Given the wavelength dependence of the relative \lya\ forest contribution shown in Figure~\ref{fig:relative}, the wavelength range over which the escape fraction is computed will determine the extent of the systematic uncertainty introduced by the $\tau_{Lya}-\delta$ relation.
Ionizing radiation from galaxies is typically measured  either through spectroscopic observations over narrow wavelength intervals  \citep[$\approx 10$~\AA][]{Bridge2010, flury_low-redshift_2022, shapley_rest-frame_2003}, or through imaging data using filters with bandwidths ranging from a few tens to several hundred \AA\ \citep{Citro2024,Siana2007,Bridge2010}. While measurements  over smaller wavelength intervals are more sensitive to the dependence of $\tau_{Ly\alpha}$ on  environmental density, we find that even using a 200\AA\ interval has a significant impact: going from overdense ($\delta = 1.5$) to underdense ($\delta = -0.5$) regions, the \lya\ forest transmission increases from 58\% to 74\%.  We expect that spectroscopic measurements of \fesc\ will be even more affected by this bias, although current data do not allow for a quantitative estimate of the effect.

Wide-field surveys, such as Euclid \citep{EuclidSkyOverview}, will provide the necessary photometric data to construct, over vast areas of the sky, galaxy overdensity maps similar to those used in our analysis. By leveraging these maps alongside our measured $\tau_{Ly\alpha}-\delta$ relation and detailed IGM absorption models at specific redshifts, it will be possible to refine IGM corrections based on QSO sight lines and better account for the influence of large-scale structure. Additionally, the bias introduced by the observed correlation between optical depth and overdensity can be mitigated by selecting survey fields that are widely separated on the sky-- at scales larger than the typical size of overdense structures ($\approx$1 degree). This approach, adopted in programs such as the Hubble Parallel Ionizing Emissivity Survey \citep{ScarlataHST,Beckett2025}, will ensure that escape fraction measurements remain unbiased by local environmental effects.


\section{Conclusions}
By adopting a photometric approach leveraging the extensive COSMOS data, we explore the \lya\ forest and its relation to cosmic density fields. We present evidence of a weak but significant correlation between galaxy overdensity and the \lya\ forest optical depth, indicating that overdense regions at $z \approx 2.5$ exhibit reduced \lya\ forest transmission. This supports the hypothesis that such regions are enriched in neutral gas, which may contribute to elevated star formation rates. Comparing our results to previous studies, including Mukae et al. (2017) and Liang et al. (2021), we find consistent trends but note a shallower correlation slope likely due to the larger spatial scales used in our analysis. 

Our findings further imply that the large-scale structure of the intergalactic medium affects the measurements of ionizing radiation escape fractions. Specifically, galaxies behind overdense regions  require larger IGM corrections compared to those behind underdense environments, underscoring the importance of accounting for the environmental density of foreground galaxies in studies of the escape fraction of ionizing radiation. To minimize this bias, we suggest using fields separated in the sky by many degrees.

\begin{acknowledgments}
We thank the referee for their valuable comments, which have helped improve the clarity of the paper. C.S. thanks Francesco Haardt for useful discussions. This research is based on observations made with the NASA/ESA Hubble Space Telescope obtained from the Space Telescope Science Institute, which is operated by the Association of Universities for Research in Astronomy, Inc., under NASA contract NAS 5--26555.
M.J.H. is supported by the Swedish Research Council (Vetenskapsr\aa{}det) and is Fellow of the Knut \& Alice Wallenberg Foundation. 
\end{acknowledgments}

\vspace{5mm}

\software{
Astropy \citep{astropycollaboration2022},
IPython \citep{perez2007},
Jupyter \citep{kluyver2016},
Matplotlib \citep{caswell2023}
Numpy \citep{harris2020},
Scipy \citep{virtanen2020},
          }


\begin{thebibliography}{}
\expandafter\ifx\csname natexlab\endcsname\relax\def\natexlab#1{#1}\fi

\bibitem[{{Adamo} {et~al.}(2024){Adamo}, {Atek}, {Bagley}, {Ba{\~n}ados},
  {Barrow}, {Berg}, {Bezanson}, {Brada{\v{c}}}, {Brammer}, {Carnall},
  {Chisholm}, {Coe}, {Dayal}, {Eisenstein}, {Eldridge}, {Ferrara}, {Fujimoto},
  {de Graaff}, {Habouzit}, {Hutchison}, {Kartaltepe}, {Kassin}, {Kriek},
  {Labb{\'e}}, {Maiolino}, {Marques-Chaves}, {Maseda}, {Mason}, {Matthee},
  {McQuinn}, {Meynet}, {Naidu}, {Oesch}, {Pentericci},
  {P{\'e}rez-Gonz{\'a}lez}, {Rigby}, {Roberts-Borsani}, {Schaerer}, {Shapley},
  {Stark}, {Stiavelli}, {Strom}, {Vanzella}, {Wang}, {Wilkins}, {Williams},
  {Willott}, {Wylezalek}, \& {Nota}}]{Adamo2024}
{Adamo}, A., {Atek}, H., {Bagley}, M.~B., {et~al.} 2024, arXiv e-prints,
  arXiv:2405.21054

\bibitem[{{Astropy Collaboration} {et~al.}(2022){Astropy Collaboration},
  {Price-Whelan}, Lim, Earl, Starkman, Bradley, Shupe, Patil, Corrales,
  Brasseur, N{\"o}the, Donath, Tollerud, Morris, Ginsburg, Vaher, Weaver,
  Tocknell, Jamieson, {van Kerkwijk}, Robitaille, Merry, Bachetti, G{\"u}nther,
  Aldcroft, {Alvarado-Montes}, Archibald, B{\'o}di, Bapat, Barentsen,
  Baz{\'a}n, Biswas, Boquien, Burke, Cara, Cara, Conroy, Conseil, Craig, Cross,
  Cruz, D'Eugenio, Dencheva, Devillepoix, Dietrich, Eigenbrot, Erben, Ferreira,
  {Foreman-Mackey}, Fox, Freij, Garg, Geda, Glattly, Gondhalekar, Gordon,
  Grant, Greenfield, Groener, Guest, Gurovich, Handberg, Hart,
  {Hatfield-Dodds}, Homeier, Hosseinzadeh, Jenness, Jones, Joseph, Kalmbach,
  Karamehmetoglu, Ka{\l}uszy{\'n}ski, Kelley, Kern, Kerzendorf, Koch, Kulumani,
  Lee, Ly, Ma, MacBride, Maljaars, Muna, Murphy, Norman, O'Steen, Oman,
  Pacifici, Pascual, {Pascual-Granado}, Patil, Perren, Pickering, Rastogi,
  Roulston, Ryan, Rykoff, Sabater, Sakurikar, Salgado, Sanghi, Saunders,
  Savchenko, Schwardt, {Seifert-Eckert}, Shih, Jain, Shukla, Sick, Simpson,
  Singanamalla, Singer, Singhal, Sinha, Sip{\H o}cz, Spitler, Stansby,
  Streicher, {\v S}umak, Swinbank, Taranu, Tewary, Tremblay, {de Val-Borro},
  Van~Kooten, Vasovi{\'c}, Verma, {de Miranda Cardoso}, Williams, Wilson,
  Winkel, {Wood-Vasey}, Xue, Yoachim, Zhang, Zonca, \& {Astropy Project
  Contributors}}]{astropycollaboration2022}
{Astropy Collaboration}, {Price-Whelan}, A.~M., Lim, P.~L., {et~al.} 2022, ApJ,
  935, 167

\bibitem[{Becker {et~al.}(2015)Becker, Bolton, Madau, Pettini, Ryan-Weber, \&
  Venemans}]{becker_evidence_2015}
Becker, G.~D., Bolton, J.~S., Madau, P., {et~al.} 2015, Monthly Notices of the
  Royal Astronomical Society, 447, 3402

\bibitem[{{Becker} {et~al.}(2021){Becker}, {D'Aloisio}, {Christenson}, {Zhu},
  {Worseck}, \& {Bolton}}]{Becker2021}
{Becker}, G.~D., {D'Aloisio}, A., {Christenson}, H.~M., {et~al.} 2021, \mnras,
  508, 1853

\bibitem[{{Becker} {et~al.}(2018){Becker}, {Davies}, {Furlanetto}, {Malkan},
  {Boera}, \& {Douglass}}]{becker2018}
{Becker}, G.~D., {Davies}, F.~B., {Furlanetto}, S.~R., {et~al.} 2018, \apj,
  863, 92

\bibitem[{{Beckett} {et~al.}(2025){Beckett}, {Rafelski}, {Scarlata}, {Hu},
  {Kim}, {Goovaerts}, {Malkan}, {Webb}, {Teplitz}, {Hayes}, {Mehta}, {Alavi},
  {Bunker}, {Citro}, {Hathi}, {Henry}, {Le Reste}, {Moretti}, {Rutkowski},
  {Trebitsch}, \& {Zanella}}]{Beckett2025}
{Beckett}, A., {Rafelski}, M., {Scarlata}, C., {et~al.} 2025, arXiv e-prints,
  arXiv:2503.20878

\bibitem[{{Bridge} {et~al.}(2010){Bridge}, {Teplitz}, {Siana}, {Scarlata},
  {Conselice}, {Ferguson}, {Brown}, {Salvato}, {Rudie}, {de Mello}, {Colbert},
  {Gardner}, {Giavalisco}, \& {Armus}}]{Bridge2010}
{Bridge}, C.~R., {Teplitz}, H.~I., {Siana}, B., {et~al.} 2010, \apj, 720, 465

\bibitem[{Bruzual \& Charlot(2003)}]{bruzual_stellar_2003}
Bruzual, G., \& Charlot, S. 2003, Monthly Notices of the Royal Astronomical
  Society, 344, 1000

\bibitem[{Calzetti {et~al.}(2000)Calzetti, Armus, Bohlin, Kinney, Koornneef, \&
  Storchi-Bergmann}]{calzetti_dust_2000}
Calzetti, D., Armus, L., Bohlin, R.~C., {et~al.} 2000, The Astrophysical
  Journal, 533, 682, publisher: IOP Publishing

\bibitem[{{Carnall} {et~al.}(2018){Carnall}, {McLure}, {Dunlop}, \&
  {Dav{\'e}}}]{Carnall2018}
{Carnall}, A.~C., {McLure}, R.~J., {Dunlop}, J.~S., \& {Dav{\'e}}, R. 2018,
  \mnras, 480, 4379

\bibitem[{Caswell {et~al.}(2023)Caswell, {Sales de Andrade}, Lee, Droettboom,
  Hoffmann, Klymak, Hunter, Firing, Stansby, Varoquaux, Hedegaard~Nielsen,
  Gustafsson, Sunden, Root, May, Elson, Sepp{\"a}nen, {hannah}, Lee, Dale,
  McDougall, Straw, Hobson, Lucas, Comer, Gohlke, Vincent, Yu, Ma, \&
  Silvester}]{caswell2023}
Caswell, T.~A., {Sales de Andrade}, E., Lee, A., {et~al.} 2023,
  Matplotlib/Matplotlib: {{REL}}: V3.7.4, ,

\bibitem[{{Citro} {et~al.}(2024){Citro}, {Scarlata}, {Mantha}, {Williams},
  {Rafelski}, {Revalski}, {Hayes}, {Henry}, {Rutkowski}, \&
  {Teplitz}}]{Citro2024}
{Citro}, A., {Scarlata}, C.~M., {Mantha}, K.~B., {et~al.} 2024, arXiv e-prints,
  arXiv:2406.07618

\bibitem[{{Euclid Collaboration: Mellier} {et~al.}(2024){Euclid Collaboration:
  Mellier}, {Abdurro'uf}, {Acevedo~Barroso}, {et~al.}}]{EuclidSkyOverview}
{Euclid Collaboration: Mellier}, Y., {Abdurro'uf}, {Acevedo~Barroso}, J.,
  {et~al.} 2024, \aap, accepted, arXiv:2405.13491

\bibitem[{Fan {et~al.}(2006)Fan, Strauss, Becker, White, Gunn, Knapp, Richards,
  Schneider, Brinkmann, \& Fukugita}]{fan_constraining_2006}
Fan, X., Strauss, M.~A., Becker, R.~H., {et~al.} 2006, The Astronomical
  Journal, 132, 117, publisher: IOP Publishing

\bibitem[{Faucher-Giguère {et~al.}(2008)Faucher-Giguère, Prochaska, Lidz,
  Hernquist, \& Zaldarriaga}]{faucher-giguere_direct_2008}
Faucher-Giguère, C.-A., Prochaska, J.~X., Lidz, A., Hernquist, L., \&
  Zaldarriaga, M. 2008, The Astrophysical Journal, 681, 831, publisher: IOP
  Publishing

\bibitem[{{Finkelstein} {et~al.}(2024){Finkelstein}, {Leung}, {Bagley},
  {Dickinson}, {Ferguson}, {Papovich}, {Akins}, {Arrabal Haro}, {Dav{\'e}},
  {Dekel}, {Kartaltepe}, {Kocevski}, {Koekemoer}, {Pirzkal}, {Somerville},
  {Yung}, {Amor{\'\i}n}, {Backhaus}, {Behroozi}, {Bisigello}, {Bromm}, {Casey},
  {Ch{\'a}vez Ortiz}, {Cheng}, {Chworowsky}, {Cleri}, {Cooper}, {Davis}, {de la
  Vega}, {Elbaz}, {Franco}, {Fontana}, {Fujimoto}, {Giavalisco}, {Grogin},
  {Holwerda}, {Huertas-Company}, {Hirschmann}, {Iyer}, {Jogee}, {Jung},
  {Larson}, {Lucas}, {Mobasher}, {Morales}, {Morley}, {Mukherjee},
  {P{\'e}rez-Gonz{\'a}lez}, {Ravindranath}, {Rodighiero}, {Rowland},
  {Tacchella}, {Taylor}, {Trump}, \& {Wilkins}}]{Finkelstein2024}
{Finkelstein}, S.~L., {Leung}, G. C.~K., {Bagley}, M.~B., {et~al.} 2024, \apjl,
  969, L2

\bibitem[{Fletcher {et~al.}(2019)Fletcher, Tang, Robertson, Nakajima, Ellis,
  Stark, \& Inoue}]{fletcher_lyman_2019}
Fletcher, T.~J., Tang, M., Robertson, B.~E., {et~al.} 2019, The Astrophysical
  Journal, 878, 87, publisher: The American Astronomical Society

\bibitem[{Flury {et~al.}(2022)Flury, Jaskot, Ferguson, Worseck, Makan,
  Chisholm, Saldana-Lopez, Schaerer, McCandliss, Xu, Wang, Oey, Ford, Heckman,
  Ji, Giavalisco, Amorín, Atek, Blaizot, Borthakur, Carr, Castellano, Barros,
  Dickinson, Finkelstein, Fleming, Fontanot, Garel, Grazian, Hayes, Henry,
  Mauerhofer, Micheva, Ostlin, Papovich, Pentericci, Ravindranath, Rosdahl,
  Rutkowski, Santini, Scarlata, Teplitz, Thuan, Trebitsch, Vanzella, \&
  Verhamme}]{flury_low-redshift_2022}
Flury, S.~R., Jaskot, A.~E., Ferguson, H.~C., {et~al.} 2022, The Astrophysical
  Journal, 930, 126, publisher: The American Astronomical Society

\bibitem[{{Gordon} {et~al.}(2023){Gordon}, {Clayton}, {Decleir}, {Fitzpatrick},
  {Massa}, {Misselt}, \& {Tollerud}}]{Gordon2023}
{Gordon}, K.~D., {Clayton}, G.~C., {Decleir}, M., {et~al.} 2023, \apj, 950, 86

\bibitem[{{Harikane} {et~al.}(2024){Harikane}, {Nakajima}, {Ouchi}, {Umeda},
  {Isobe}, {Ono}, {Xu}, \& {Zhang}}]{Harikane2024}
{Harikane}, Y., {Nakajima}, K., {Ouchi}, M., {et~al.} 2024, \apj, 960, 56

\bibitem[{Harris {et~al.}(2020)Harris, Millman, {van der Walt}, Gommers,
  Virtanen, Cournapeau, Wieser, Taylor, Berg, Smith, Kern, Picus, Hoyer, {van
  Kerkwijk}, Brett, Haldane, {del R{\'i}o}, Wiebe, Peterson,
  {G{\'e}rard-Marchant}, Sheppard, Reddy, Weckesser, Abbasi, Gohlke, \&
  Oliphant}]{harris2020}
Harris, C.~R., Millman, K.~J., {van der Walt}, S.~J., {et~al.} 2020, Nature,
  585, 357

\bibitem[{Hayes {et~al.}(2021)Hayes, Runnholm, Gronke, \&
  Scarlata}]{hayes_spectral_2021}
Hayes, M.~J., Runnholm, A., Gronke, M., \& Scarlata, C. 2021, The Astrophysical
  Journal, 908, 36, publisher: The American Astronomical Society

\bibitem[{{Hayes} {et~al.}(2024){Hayes}, {Saldana-Lopez}, {Citro}, {James},
  {Mingozzi}, {Scarlata}, {Martinez}, \& {Berg}}]{Hayes2024}
{Hayes}, M.~J., {Saldana-Lopez}, A., {Citro}, A., {et~al.} 2024, arXiv
  e-prints, arXiv:2411.09262

\bibitem[{{Inoue} {et~al.}(2014){Inoue}, {Shimizu}, {Iwata}, \&
  {Tanaka}}]{Inoue2014}
{Inoue}, A.~K., {Shimizu}, I., {Iwata}, I., \& {Tanaka}, M. 2014, \mnras, 442,
  1805

\bibitem[{Inoue {et~al.}(2014)Inoue, Shimizu, Iwata, \&
  Tanaka}]{inoue_updated_2014}
Inoue, A.~K., Shimizu, I., Iwata, I., \& Tanaka, M. 2014, Monthly Notices of
  the Royal Astronomical Society, 442, 1805

\bibitem[{Khostovan {et~al.}(2025)Khostovan, Kartaltepe, Salvato, Ilbert,
  Casey, Algera, Antwi-Danso, Battisti, Brinch, Brusa, Calabro, Capak, Chartab,
  Cooper, Cox, Darvish, Drakos, Faisst, George, Gozaliasl, Harish, Hasinger,
  Hatamnia, Iovino, Jin, Kashino, Koekemoer, Laishram, Lee, Lertprasertpong,
  Lilly, Masters, Mobasher, Nagao, Onodera, Peng, Sanders, Sanders, Sattari,
  Scoville, Shah, Silverman, Suzuki, Tanaka, Toft, Trakhtenbrot, Trump,
  Vaccari, Valentino, Vanderhoof, Weaver, Yun, \& Zavala}]{cosmos_z}
Khostovan, A.~A., Kartaltepe, J.~S., Salvato, M., {et~al.} 2025, COSMOS
  Spectroscopic Redshift Compilation (First Data Release): 165k Redshifts
  Encompassing Two Decades of Spectroscopy, , , arXiv:2503.00120

\bibitem[{Kirkman {et~al.}(2005)Kirkman, Tytler, Suzuki, Melis, Hollywood,
  James, So, Lubin, Jena, Norman, \& Paschos}]{kirkman_h_2005}
Kirkman, D., Tytler, D., Suzuki, N., {et~al.} 2005, Monthly Notices of the
  Royal Astronomical Society, 360, 1373

\bibitem[{Kluyver {et~al.}(2016)Kluyver, {Ragan-Kelley}, P{\'e}rez, Granger,
  Bussonnier, Frederic, Kelley, Hamrick, Grout, Corlay, Ivanov, Avila, Abdalla,
  Willing, \& {Jupyter Development Team}}]{kluyver2016}
Kluyver, T., {Ragan-Kelley}, B., P{\'e}rez, F., {et~al.} 2016, Jupyter
  {{Notebooks}}---a Publishing Format for Reproducible Computational Workflows,
  87--90

\bibitem[{Kroupa(2001)}]{kroupa_variation_2001}
Kroupa, P. 2001, Monthly Notices of the Royal Astronomical Society, 322, 231

\bibitem[{{Laloux} {et~al.}(2023){Laloux}, {Georgakakis}, {Andonie},
  {Alexander}, {Ruiz}, {Rosario}, {Aird}, {Buchner}, {Carrera}, {Lapi}, {Ramos
  Almeida}, {Salvato}, \& {Shankar}}]{Laloux2023}
{Laloux}, B., {Georgakakis}, A., {Andonie}, C., {et~al.} 2023, \mnras, 518,
  2546

\bibitem[{Lemaux {et~al.}(2022)Lemaux, Cucciati, Fèvre, Zamorani, Lubin,
  Hathi, Ilbert, Pelliccia, Amorín, Bardelli, Cassata, Gal, Garilli, Guaita,
  Giavalisco, Hung, Koekemoer, Maccagni, Pentericci, Ribeiro, Schaerer, Shah,
  Shen, Staab, Talia, Thomas, Tomczak, Tresse, Vanzella, Vergani, \&
  Zucca}]{lemaux_vimos_2022}
Lemaux, B.~C., Cucciati, O., Fèvre, O.~L., {et~al.} 2022, Astronomy \&
  Astrophysics, 662, A33, publisher: EDP Sciences

\bibitem[{{Liang} {et~al.}(2021){Liang}, {Kashikawa}, {Cai}, {Fan},
  {Prochaska}, {Shimasaku}, {Tanaka}, {Uchiyama}, {Ito}, {Shimakawa},
  {Nagamine}, {Shimizu}, {Onoue}, \& {Toshikawa}}]{Liang2021}
{Liang}, Y., {Kashikawa}, N., {Cai}, Z., {et~al.} 2021, \apj, 907, 3

\bibitem[{Lin {et~al.}(2024)Lin, Scarlata, Williams, Chen, Kelly, Langeroodi,
  Hjorth, Chisholm, Koekemoer, Zitrin, \& Diego}]{lin_empirical_2024}
Lin, Y.-H., Scarlata, C., Williams, H., {et~al.} 2024, Monthly Notices of the
  Royal Astronomical Society, 527, 4173

\bibitem[{{Liske} {et~al.}(2000){Liske}, {Webb}, {Williger},
  {Fern{\'a}ndez-Soto}, \& {Carswell}}]{Liske2000}
{Liske}, J., {Webb}, J.~K., {Williger}, G.~M., {Fern{\'a}ndez-Soto}, A., \&
  {Carswell}, R.~F. 2000, \mnras, 311, 657

\bibitem[{{Madau}(1995)}]{Madau1995}
{Madau}, P. 1995, \apj, 441, 18

\bibitem[{{Madau} {et~al.}(2024){Madau}, {Giallongo}, {Grazian}, \&
  {Haardt}}]{Madau2024}
{Madau}, P., {Giallongo}, E., {Grazian}, A., \& {Haardt}, F. 2024, \apj, 971,
  75

\bibitem[{{Mawatari} {et~al.}(2023){Mawatari}, {Inoue}, {Yamada}, {Hayashino},
  {Prochaska}, {Lee}, {Tejos}, {Kashikawa}, {Otsuka}, {Yamanaka}, {Schlegel},
  {Matsuda}, {Hennawi}, {Iwata}, {Umehata}, {Mukae}, {Ouchi}, {Sugahara}, \&
  {Tamura}}]{Mawatari2023}
{Mawatari}, K., {Inoue}, A.~K., {Yamada}, T., {et~al.} 2023, \aj, 165, 208

\bibitem[{Monzon {et~al.}(2020)Monzon, Prochaska, Lee, \&
  Chisholm}]{monzon_effective_2020}
Monzon, J.~S., Prochaska, J.~X., Lee, K.-G., \& Chisholm, J. 2020, The
  Astronomical Journal, 160, 37

\bibitem[{Mukae {et~al.}(2017)Mukae, Ouchi, Kakiichi, Suzuki, Ono, Cai, Inoue,
  Chiang, Shibuya, \& Matsuda}]{mukae_cosmic_2017}
Mukae, S., Ouchi, M., Kakiichi, K., {et~al.} 2017, The Astrophysical Journal,
  835, 281, publisher: The American Astronomical Society

\bibitem[{{Naidu} {et~al.}(2017){Naidu}, {Oesch}, {Reddy}, {Holden}, {Steidel},
  {Montes}, {Atek}, {Bouwens}, {Carollo}, {Cibinel}, {Illingworth},
  {Labb{\'e}}, {Magee}, {Morselli}, {Nelson}, {van Dokkum}, \&
  {Wilkins}}]{Naidu2017}
{Naidu}, R.~P., {Oesch}, P.~A., {Reddy}, N., {et~al.} 2017, \apj, 847, 12

\bibitem[{{Oke}(1990)}]{Oke1990}
{Oke}, J.~B. 1990, \aj, 99, 1621

\bibitem[{Perez \& Granger(2007)}]{perez2007}
Perez, F., \& Granger, B.~E. 2007, Computing in Science and Engineering, 9, 21

\bibitem[{{Prochaska}(2019)}]{Prochaska2019}
{Prochaska}, J.~X. 2019, Saas-Fee Advanced Course, 46, 111

\bibitem[{Reddy {et~al.}(2016)Reddy, Steidel, Pettini, Bogosavljević, \&
  Shapley}]{reddy_connection_2016}
Reddy, N.~A., Steidel, C.~C., Pettini, M., Bogosavljević, M., \& Shapley,
  A.~E. 2016, The Astrophysical Journal, 828, 108, publisher: The American
  Astronomical Society

\bibitem[{Rudie {et~al.}(2013)Rudie, Steidel, Shapley, \&
  Pettini}]{rudie_column_2013}
Rudie, G.~C., Steidel, C.~C., Shapley, A.~E., \& Pettini, M. 2013, The
  Astrophysical Journal, 769, 146, publisher: The American Astronomical Society

\bibitem[{{Saxena} {et~al.}(2022){Saxena}, {Pentericci}, {Ellis}, {Guaita},
  {Calabr{\`o}}, {Schaerer}, {Vanzella}, {Amor{\'\i}n}, {Bolzonella},
  {Castellano}, {Fontanot}, {Hathi}, {Hibon}, {Llerena}, {Mannucci},
  {Saldana-Lopez}, {Talia}, \& {Zamorani}}]{Saxena2022}
{Saxena}, A., {Pentericci}, L., {Ellis}, R.~S., {et~al.} 2022, \mnras, 511, 120

\bibitem[{{Scarlata} {et~al.}(2022){Scarlata}, {Alavi}, {Bruton}, {Bunker},
  {Colbert}, {Grazian}, {Haardt}, {Hathi}, {Hayes}, {Henry}, {Malkan}, {Mehta},
  {Moretti}, {Rafelski}, {Rutkowski}, {Teplitz}, {Trebitsch}, \&
  {Zanella}}]{ScarlataHST}
{Scarlata}, C., {Alavi}, A., {Bruton}, S.~T., {et~al.} 2022, {The Parallel
  Ionizing Emissivity Survey}, HST Proposal. Cycle 30, ID. \#17147, ,

\bibitem[{{Scoville} {et~al.}(2007){Scoville}, {Aussel}, {Brusa}, {Capak},
  {Carollo}, {Elvis}, {Giavalisco}, {Guzzo}, {Hasinger}, {Impey}, {Kneib},
  {LeFevre}, {Lilly}, {Mobasher}, {Renzini}, {Rich}, {Sanders}, {Schinnerer},
  {Schminovich}, {Shopbell}, {Taniguchi}, \& {Tyson}}]{Scoville2007}
{Scoville}, N., {Aussel}, H., {Brusa}, M., {et~al.} 2007, \apjs, 172, 1

\bibitem[{Shapley {et~al.}(2003)Shapley, Steidel, Pettini, \&
  Adelberger}]{shapley_rest-frame_2003}
Shapley, A.~E., Steidel, C.~C., Pettini, M., \& Adelberger, K.~L. 2003, The
  Astrophysical Journal, 588, 65, publisher: IOP Publishing

\bibitem[{{Shi} {et~al.}(2019){Shi}, {Huang}, {Lee}, {Toshikawa}, {Bowen},
  {Malavasi}, {Lemaux}, {Cucciati}, {Le Fevre}, \& {Dey}}]{Shi2019}
{Shi}, K., {Huang}, Y., {Lee}, K.-S., {et~al.} 2019, \apj, 879, 9

\bibitem[{{Siana} {et~al.}(2007){Siana}, {Teplitz}, {Colbert}, {Ferguson},
  {Dickinson}, {Brown}, {Conselice}, {de Mello}, {Gardner}, {Giavalisco}, \&
  {Menanteau}}]{Siana2007}
{Siana}, B., {Teplitz}, H.~I., {Colbert}, J., {et~al.} 2007, \apj, 668, 62

\bibitem[{{Siana} {et~al.}(2015){Siana}, {Shapley}, {Kulas}, {Nestor},
  {Steidel}, {Teplitz}, {Alavi}, {Brown}, {Conselice}, {Ferguson}, {Dickinson},
  {Giavalisco}, {Colbert}, {Bridge}, {Gardner}, \& {de Mello}}]{Siana2015}
{Siana}, B., {Shapley}, A.~E., {Kulas}, K.~R., {et~al.} 2015, \apj, 804, 17

\bibitem[{{Slosar} {et~al.}(2011){Slosar}, {Font-Ribera}, {Pieri}, {Rich}, {Le
  Goff}, {Aubourg}, {Brinkmann}, {Busca}, {Carithers}, {Charlassier},
  {Cort{\^e}s}, {Croft}, {Dawson}, {Eisenstein}, {Hamilton}, {Ho}, {Lee},
  {Lupton}, {McDonald}, {Medolin}, {Muna}, {Miralda-Escud{\'e}}, {Myers},
  {Nichol}, {Palanque-Delabrouille}, {P{\^a}ris}, {Petitjean}, {Pi{\v{s}}kur},
  {Rollinde}, {Ross}, {Schlegel}, {Schneider}, {Sheldon}, {Weaver}, {Weinberg},
  {Yeche}, \& {York}}]{Slosar2011}
{Slosar}, A., {Font-Ribera}, A., {Pieri}, M.~M., {et~al.} 2011, \jcap, 2011,
  001

\bibitem[{{Steidel} {et~al.}(2018){Steidel}, {Bogosavljevi{\'c}}, {Shapley},
  {Reddy}, {Rudie}, {Pettini}, {Trainor}, \& {Strom}}]{Steidel2018}
{Steidel}, C.~C., {Bogosavljevi{\'c}}, M., {Shapley}, A.~E., {et~al.} 2018,
  \apj, 869, 123

\bibitem[{Steidel {et~al.}(2018)Steidel, Bogosavljević, Shapley, Reddy, Rudie,
  Pettini, Trainor, \& Strom}]{steidel_keck_2018}
Steidel, C.~C., Bogosavljević, M., Shapley, A.~E., {et~al.} 2018, The
  Astrophysical Journal, 869, 123, publisher: The American Astronomical Society

\bibitem[{Taamoli {et~al.}(2024)Taamoli, Mobasher, Chartab, Darvish, Weaver,
  Hemmati, Casey, Sattari, Brammer, Capak, Ilbert, Kartaltepe, McCracken,
  Moneti, Sanders, Scoville, Steinhardt, \& Toft}]{taamoli_large-scale_2024}
Taamoli, S., Mobasher, B., Chartab, N., {et~al.} 2024, The Astrophysical
  Journal, 966, 18, publisher: The American Astronomical Society

\bibitem[{{Toshikawa} {et~al.}(2016){Toshikawa}, {Kashikawa}, {Overzier},
  {Malkan}, {Furusawa}, {Ishikawa}, {Onoue}, {Ota}, {Tanaka}, {Niino}, \&
  {Uchiyama}}]{Toshikawa2016}
{Toshikawa}, J., {Kashikawa}, N., {Overzier}, R., {et~al.} 2016, \apj, 826, 114

\bibitem[{{Vanzella} {et~al.}(2018){Vanzella}, {Nonino}, {Cupani},
  {Castellano}, {Sani}, {Mignoli}, {Calura}, {Meneghetti}, {Gilli}, {Comastri},
  {Mercurio}, {Caminha}, {Caputi}, {Rosati}, {Grillo}, {Cristiani}, {Balestra},
  {Fontana}, \& {Giavalisco}}]{Vanzella2018}
{Vanzella}, E., {Nonino}, M., {Cupani}, G., {et~al.} 2018, \mnras, 476, L15

\bibitem[{Virtanen {et~al.}(2020)Virtanen, Gommers, Oliphant, Haberland, Reddy,
  Cournapeau, Burovski, Peterson, Weckesser, Bright, {van der Walt}, Brett,
  Wilson, Millman, Mayorov, Nelson, Jones, Kern, Larson, Carey, Polat, Feng,
  Moore, VanderPlas, Laxalde, Perktold, Cimrman, Henriksen, Quintero, Harris,
  Archibald, Ribeiro, Pedregosa, {van Mulbregt}, \& {SciPy 1. 0
  Contributors}}]{virtanen2020}
Virtanen, P., Gommers, R., Oliphant, T.~E., {et~al.} 2020, Nature Methods, 17,
  261

\bibitem[{Weaver {et~al.}(2022)Weaver, Kauffmann, Ilbert, McCracken, Moneti,
  Toft, Brammer, Shuntov, Davidzon, Hsieh, Laigle, Anastasiou, Jespersen,
  Vinther, Capak, Casey, McPartland, Milvang-Jensen, Mobasher, Sanders,
  Zalesky, Arnouts, Aussel, Dunlop, Faisst, Franx, Furtak, Fynbo, Gould, Greve,
  Gwyn, Kartaltepe, Kashino, Koekemoer, Kokorev, Fèvre, Lilly, Masters,
  Magdis, Mehta, Peng, Riechers, Salvato, Sawicki, Scarlata, Scoville, Shirley,
  Silverman, Sneppen, Smolc̆ić, Steinhardt, Stern, Tanaka, Taniguchi,
  Teplitz, Vaccari, Wang, \& Zamorani}]{weaver_cosmos2020_2022}
Weaver, J.~R., Kauffmann, O.~B., Ilbert, O., {et~al.} 2022, The Astrophysical
  Journal Supplement Series, 258, 11, publisher: The American Astronomical
  Society

\bibitem[{{Williams} {et~al.}(2023){Williams}, {Kelly}, {Chen}, {Brammer},
  {Zitrin}, {Treu}, {Scarlata}, {Koekemoer}, {Oguri}, {Lin}, {Diego}, {Nonino},
  {Hjorth}, {Langeroodi}, {Broadhurst}, {Rogers}, {Perez-Fournon}, {Foley},
  {Jha}, {Filippenko}, {Strolger}, {Pierel}, {Poidevin}, \&
  {Yang}}]{Williams2023}
{Williams}, H., {Kelly}, P.~L., {Chen}, W., {et~al.} 2023, Science, 380, 416

\end{thebibliography}
\end{document}